\documentclass[prl,twocolumn,preprintnumbers,nofootinbib,showpacs,amsmath,amssymb,superscriptaddress]{revtex4-1}
\usepackage{psfrag,graphicx}

\usepackage[
      colorlinks=true,
      linkcolor=blue,
      urlcolor=blue,
      filecolor=black,
      citecolor=blue,
            ]{hyperref}
\usepackage{color}
\usepackage{amsfonts}
\usepackage{amssymb}
\usepackage{epstopdf}
\usepackage{dsfont}
\usepackage{epsfig}

\begin{document}
\title{Scaling Laws in Gravitational Collapse}
\author{Rong-Gen Cai}
\email{cairg@itp.ac.cn}

\author{Run-Qiu Yang}
\email{aqiu@itp.ac.cn}
\affiliation{Institute of Theoretical Physics,
 Chinese Academy of Sciences,Beijing 100190, China.}

\begin{abstract}
This paper presents  two interesting scaling laws, which relate some critical exponents in the critical behavior of spherically symmetric gravitational collapses. These scaling laws are independent of the details of gravity theory under consideration and share similar forms as those in thermodynamic and geometrical phase transitions in condensed matter system.  The properties of the scaling laws are discussed and  some numerical checks are given.

\end{abstract}
\maketitle

\noindent

\section{Introduction}
\label{Introd}
Scaling laws, critical exponents and universal classes are the central concepts  for phase transition in thermodynamic systems and complex systems.  While the critical exponents
characterize the critical behavior of the systems near critical point, the scaling laws unify  the critical behaviors in various different systems into a few universal relations. In 1993,   the  pioneering work by  Choptuik~\cite{Choptuik} showed that a named type II critical behavior exists in gravitational collapse of a massless scalar field in asymptotically flat space-time.
 By numerical simulation, Choptuik found that there is a critical case $p=p^*$, where $p$ is a parameter which characterizes the amplitude of the initial configuration and $p^*$ is the  critical value where  black hole can form, and that in the case $p\rightarrow p^{*+}$, a power-law form of the black-hole mass appears,
\begin{equation}\label{massp1}
    M_h\propto(p-p^*)^\beta, ~~~p\rightarrow p^{*+}.
\end{equation}
It was found that  $\beta\simeq0.37$, and it is universal for a large class of  initial  configurations in gravitational collapse. This scaling relation with the self-similarity  at the critical solution has drawn a  lot of attention since then. See Ref. \cite{Gundlach:2007gc} for a recent review.

In a more general system, by choose different tuning parameter or models, one can obtain different critical exponents. For example, in the case of  gravitational collapse of charged scalar field, one can take the charge $e$  of the scalar field as the tuning parameter. In that case the critical exponent in the mass scaling relation with respect  to $e$ is approximately  0.74~\cite{Hod:1996ar}.  In the perfect fluid collapse, the exponent $\beta$ strongly depends on the value of $k$ in the equation of state $P=k\rho$, where $P$ and $\rho$ are the pressure and energy density of fluid \cite{Maison:1995cc}.

Over the past years, the  gravitational collapse in an closed system has been studied intensively due to the discovery of  the so called weakly turbulence in anti-de Sitter (AdS) space~\cite{Bizon:2011gg} or in  a cavity  with perfect reflection wall at a finite distance from the origin~ \cite{Maliborski:2012gx}. In such systems, there exist  infinite critical solutions in principle. According to the very recent work~\cite{Olivan:2015fmy}, for every critical solution, there are at least two different  mass scaling relations: one is the same as the one in (\ref{massp1}) with critical exponent $0.37$, the other has a critical exponent $\xi \simeq 0.7$ with a mass gap.
One may expect that, in such systems, if choose different quantity in  theories as the tuning parameter, one can obtain many different critical exponents.

Once those critical exponents appear, one may naturally ask wether there exist some scaling laws to relate them, like the scaling laws in thermodynamic systems? As the choice of the tuning parameter is of some arbitrariness (such as the parameters in initial value families, the parameters in the theory under consideration), can we classify them into a few classes? In this paper, we will present two universal scaling laws in the spherically symmetrical gravitational collapse.  They build a bridge for scaling relations between different tuning parameters and theoretical models.  These scaling laws are independent of the details of the theory and also appear in other systems such as thermodynamic or geometrical phase transitions. We will discuss its properties and make some numerical checks.

\section{ critical exponents}
\label{app1}
Let us  consider a generic spherically symmetric  gravitational collapse system under a given initial value family parameterized by $p$. Besides the Newton gravitational constant $G$ for gravity, suppose that there is an additional parameter $\lambda$ in the system, which may be a parameter in theory or initial data. Here we do not make any additional assumptions on the  dynamics of gravity and matter fields, and do not  specify any special boundary conditions, except that the space-time should be spherically symmetric  in the process of gravitational collapse and there is at least one critical value for $p$ such that  $M_h=0$.
Then  we can define a group of scaling relations about the black hole mass $M_h$, tuning parameter $p$ in the initial family, the time $t$ when black hole just forms and other parameter $\lambda$ as,
\begin{equation}\label{scaling1}
\begin{split}
    &M_h|_{\lambda=\lambda0}=a^+_n|p- p_{n0}|^{\beta^+_n}, ~ p\rightarrow p_{n0}^+,\\
    &M_h|_{p= p_{n0}^+}=b^+_n|\lambda-\lambda_0|^{2-\alpha^+_n}, ~\lambda\rightarrow\lambda_0^+~\text{or}~\lambda_0^-,\\
    &\chi_{p_n}=(\partial p/\partial\lambda)|_{M_h=0}=c_n|\lambda-\lambda_0|^{\delta_n},~\lambda\rightarrow\lambda_0,\\
    &\chi_{tn}^+=(\partial t/\partial p)_{\lambda=\lambda0}=d_n^+|p-p_{n0}|^{\nu_n^+-1},~p\rightarrow p_{n0}^+.
    \end{split}
\end{equation}
Here indices $n=0,1,2,3,\cdots$, which label the different critical solutions, which appear in a closed system mentioned above. $a^+_n,b^+_n, d_n^\pm$ and $c_n$ are four proportionality coefficients, which may depend on the value of $\lambda_0$ and the initial value function families. $\lambda_0$ is a  fixed value of $\lambda$. $p_{n0}$ is the $n$-th critical amplitude when $\lambda=\lambda_0$. Here we add  $+$ to the indices of  $\beta_n, \alpha_n$ and $\nu_n$ to distinguish the different cases that $p\rightarrow p_{n0}^+$ and $p\rightarrow p_{n0}^-$. In an open system such as  the gravitational collapse in asymptotic flat space or de Sitter (dS) space-time, there is only one critical value $p$ for a given initial value family. In such a case, the mass scaling behavior can only occur when $p\rightarrow p^{*+}$. However, in the closed system such as the gravitational collapse in asymptotic anti-de Sitter(AdS) or asymptotic flat space with a perfectly reflection mirror at a fixed radial position, there are infinite critical solutions. In those cases, there exists a mass gap $M_{gn}$ when $p\rightarrow p_{n0}^-$ \cite{Olivan:2015fmy}. Then we can still define a group of critical exponents in a manner as,
\begin{equation}\label{scaling1b}
\begin{split}
    &\Delta M_h|_{\lambda=\lambda0}=a^-_n|p- p_{n0}|^{\beta^-_n}, ~~ p\rightarrow p_{n0}^-,\\
    &\Delta M_h|_{p= p_{n0}^+}=b^-_n|\lambda-\lambda_0|^{2-\alpha^-_n},\lambda\rightarrow\lambda_0^+~\text{or}~\lambda_0^-,\\
    &\chi_{tn}^-=d_n^-|p-p_{n0}|^{\nu_n^--1},~p\rightarrow p_{n0}^-.
    \end{split}
\end{equation}
where $\Delta M_h=M_h-M_{gn}$, $a^-_n$ , $b^-_n$ and $d_n^-$  are  proportionality coefficients. The scaling relations \eqref{scaling1} and \eqref{scaling1b} with  seven critical exponents $\{\beta^\pm_n, \alpha^\pm_n, \nu_n^\pm,\delta_n\}$ give out the critical behavior when $p\rightarrow p_{n0}^\pm$ and $\lambda\rightarrow\lambda_0$.\footnote{If $\chi_{p_n}>0$, the limit in the second equation of Eqs.\eqref{scaling1} is taken as $\lambda\rightarrow\lambda_0^-$. Otherwise, the limit  is done as  $\lambda\rightarrow\lambda_0^+$. The limits in the second equations in Eqs.(\ref{scaling1}) and \eqref{scaling1b} are always opposite. } The critical exponents $\beta^\pm_n$ and $\alpha^\pm_n$ describe the dependence of critical behaviors on two parameters $p$ and $\lambda$, respectively. The third and fourth relations describe how the critical parameter $p_n$ in the initial value family and the time of black hole just forming depend on $\lambda$ and $p$. These scaling relations were first introduced in Ref.~\cite{Cai:2015jbs} to describe the influence of $\lambda\phi^4$ on massless scalar collapse in AdS space-time.  Note that there is a difference in  the definition of $\chi_{tn}^\pm$ for the purpose in this paper.

\section{scaling laws}
\label{app2}
In what follows, we will take \eqref{scaling1} as an example, the case for \eqref{scaling1b} can be obtained by replacement $M_h\rightarrow M_h-M_{gn}$. Near the critical point, for a set of given $\lambda$ and $p$, the value of $M_h$ and the forming time $t_n$ can be  determined as $M_h=M_h(p, \lambda)$ and $t_n=t_n(p, \lambda_0)$. Let $\delta p=p- p_{n0}, \delta t_n=t_n(p, \lambda_0)-t_n(p_{n0}, \lambda_0)$ and $\delta\lambda=\lambda-\lambda_0\geq0$. Near the $n$-th critical solution, the value of $\delta p, \delta t_n, M_h$ and $\delta\lambda$ can be expressed as the following relations,
\begin{equation}\label{deltaep1}
    \delta p=\delta p(M_h,\delta\lambda),
\end{equation}
and,
\begin{equation}\label{deltaep2}
    \delta t=\delta t(M_h,\delta\lambda).
\end{equation}
As only two in  $\delta p, \delta t_n, M_h$ and $\delta\lambda$ are independent, we assume there exist two homogeneous functions as,
\begin{equation}\label{homo1}
   \kappa\delta p(M_h,\delta\lambda)=\delta p(\kappa^xM_h,\kappa^y\delta\lambda),
\end{equation}
and
\begin{equation}\label{homo1t}
   \kappa^{x/(D-2+z_n^+)}\delta t(\delta M_h,\delta\lambda)=\delta t(\kappa^x\delta M_h,\kappa^y\delta\lambda),
\end{equation}
where $x, y$ and $z_n^+$ are three constants and $D$ is the spatial dimensions. In the $D+1$-dimensional spherically symmetric spacetime, the mass $M_h$ has dimension [length]$^{D-2}$ and time has dimension [lenght], so $z_n^+$ can be treated as an anomalous dimension. Combine \eqref{homo1} and the first one in \eqref{scaling1} and let $\delta\lambda=0$, we can find
\begin{equation}\label{gammaxy}
    x=\beta^+_n.
\end{equation}
Furthermore, make a derivative for \eqref{homo1} with respect to $\delta\lambda$ and put it back into the definition of $\chi_{ p_n}$, we have,
\begin{equation}\label{homo2}
    \chi_{ p_n}(\delta\lambda)=\kappa^{y-1}\chi_{ p_n}(\kappa^y\delta\lambda).
\end{equation}
Take $\kappa=\delta\lambda^{-1/y}$, we have $\chi_{ p_n}(\delta\lambda)=\delta\lambda^{(1-y)/y}\chi_{ p_n}(1)$, which implies,
\begin{equation}\label{betaxy}
    \delta_n=(1-y)/y\Rightarrow y=1/(\delta_n+1).
\end{equation}
When we fix $ p= p_{n0}$, i.e., $\delta p=0$,  Eq.~\eqref{homo1} implies $\delta p(\kappa^xM_h,\kappa^y\delta\lambda)=0$. Take $\kappa=\delta\lambda^{-1/y}$, we have, $\delta p(M_h\delta\lambda^{-x/y},1)=0$, which implies that $M_h\propto \delta\lambda^{x/y}$. As a result, we have,
\begin{equation}\label{Mhxy}
    2-\alpha^+_n=x/y.
\end{equation}
Fixing $\delta\lambda=0$ and taking $\kappa=M_h^{-1/x}$, Eq. \eqref{homo1t} becomes,
\begin{equation}\label{homo2t}
\begin{split}
    \delta t(M_h)&=M_h^{1/(D-2+z_n^+)}\delta t(1)\\
    &=\delta p^{x/(D-2+z_n^+)}\delta t(1),
    \end{split}
\end{equation}
so
\begin{equation}\label{etaxy}
    \nu_n=x/(D-2+z_n^+).
\end{equation}

Now combine the results \eqref{gammaxy}, \eqref{betaxy}, \eqref{Mhxy} and \eqref{etaxy}, we find two scaling laws
\begin{equation}\label{scalingrel}
    \alpha^\pm_n+\beta^\pm_n(\delta_n+1)=2,~~\beta^\pm_n=\nu_n^\pm(D-2+z_n^\pm),
\end{equation}
which relate these critical exponents defined in (\ref{scaling1}) and (\ref{scaling1b}).
Because we did not specify any  gravitational theory, matter field in space-time, and initial value configuration,  the relations \eqref{scalingrel} are universal and should hold  in many different systems.

If we know more about the model in the gravitational collapse, the number of independent critical exponents can be further reduced. For a given $\lambda$, if we treat the expansion $\theta$ at the original point is a function of $p$, then finding critical solution forms an eigenvalue problem, i.e., find suitable $p_n$ such that $\theta|_{r=0}=0$. If the equations of motion for matter field and metric are all smoothly dependent on $\lambda$, we can assume that the eigenvalue $p_n$ is a function of $\delta\lambda$ and there is a Taylor's expansion for $p_n$ as,
\begin{equation}\label{epsilonn}
    p_n=p_{n0}+\sum_{i=1}^\infty p_{n0}^{(i)} (\delta\lambda)^i,  \  \ \ \text{as}~\delta\lambda\rightarrow0.
\end{equation}
Here $p_{n0}^{(i)}, i=1,2,3,\cdots$ are all independent of $\delta\lambda$. Let $p_{n0}^{(k)}$ is the lowest order nonzero coefficient in the expansion \eqref{epsilonn}, then put \eqref{epsilonn} into the third one in  \eqref{scaling1}, we can obtain $\delta_n=k-1$. The value $k$ is determined by the maximal  asymptotic symmetric group for $\delta\lambda$ in the case $\delta\lambda\rightarrow0$. For example, if in a specified theory there is a symmetry such that $\delta\lambda\rightarrow-\delta\lambda$, i.e., the maximal asymptotic symmetric group is $\mathds{Z}_2$ group, then the first nonzero coefficient is $p_{n0}^{(2)}$, thus we have $\delta_n=1$. For a general case, if the maximal asymptotic symmetric group for $\delta\lambda$ in the neibourhood of $\lambda=\lambda_0$ is $\mathds{Z}_m$, then $\delta_n=m-1$.

\section{Coefficient equation}
\label{property}
In the scaling relations \eqref{scaling1} and \eqref{scaling1b}, we did not assume the critical exponents are independent of the initial value family. However, if we assume they are independent of the initial value family ``locally", then the coefficients defined in \eqref{scaling1} and \eqref{scaling1b} satisfy,
\begin{equation}\label{abc0}
    a^\pm_n|c_n|^{\beta^\pm_n}/b^\pm_n=1.
\end{equation}
More exactly, for given initial data function family $f_p$\footnote{$f_p$ stands for  2-component function familiy if one needs two initial value functions such as the field configuration and  corresponding canonical momentum.}, {\it if there is a two-parameter function family $F_{p,\lambda}$ and an open internal $U \subset \mathds{R}$ such that (1) $F_{p,\lambda_0}=f_p$ and $\lambda_0\in U$; (2)$\forall \lambda\in U$, using $F_{p,\lambda}$ as the initial data function family,  all critical exponents are independent of
$\lambda$}, then  Eq. \eqref{abc0}  holds.

\begin{figure}
\begin{center}
\includegraphics[width=0.45\textwidth]{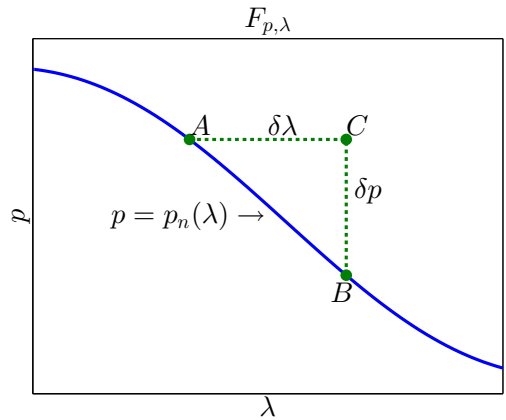}
\caption{The schematic diagram for the curve $p=p_n(\lambda)$ in the $\lambda-p$ plane. Every point in this plan corresponds to a initial data function. }
\label{plambda1}
\end{center}
\end{figure}

The proof is as follows. Let  $\beta^+_n(\lambda_0), \alpha^+_n(\lambda_0)$ and $\delta_n(\lambda_0)$ be the $n$-th critical exponents for initial function family $F_{p,\lambda_0}$ and $p_n(\lambda_0)$ is the $n$-th critical value of $p$. In Fig.~\ref{plambda1}, we plot a schematic diagram for the curve $p=p_n(\lambda)$ in the $\lambda-p$ plane. Every point in this plane corresponds to an initial value function. Let point $A$ correspond to the critical initial function at $\lambda=\lambda_0$. Then we make an infinitesimal shift on $\lambda$ such that $\lambda\rightarrow\lambda+\delta\lambda$, which causes an infinitesimal shift on the critical value of $p$ such that $p_n\rightarrow p_n+\delta p$. The new critical initial function then is denoted by $B$ in the $\lambda-p$ plane. As shown in Fig.~\ref{plambda1}, we take a point $C$ as the initial function for the gravitational collapse with $\delta p>0$.\footnote{This can always be achieved by choosing $C$ properly.}  As the point $C$ is in the neighborhood of $B$ with $\lambda=\lambda_0+\delta\lambda$ and
according to the first scaling relation in Eqs. \eqref{scaling1}, we obtain the black hole mass at point $C$ as,
\begin{equation}\label{massC1}
M_h(C)=a^+_n(\lambda_0+\delta\lambda)\delta p^{\beta_n^+(\lambda_0+\delta\lambda)}.
\end{equation}
%
Note that  the point $C$ is also in the neighborhood of $A$ with $p=p_n(\lambda_0)$, according to the second scaling relation in Eqs. \eqref{scaling1}, we can also express the  black hole mass at point $C$ as,
\begin{equation}\label{massC2}
    M_h(C)=b^+_n(\lambda_0)\delta\lambda^{2-\alpha_n^+(\lambda_0)}.
\end{equation}
%
Compare Eqs. \eqref{massC1} and \eqref{massC2},  we obtain,
\begin{equation}\label{massC3}
    \delta\lambda^{2-\alpha_n^+(\lambda_0)}=\frac{a^+_n(\lambda_0+\delta\lambda)}{b^+_n(\lambda_0)}\delta p^{\beta_n^+(\lambda_0+\delta\lambda)}.
\end{equation}
Based on the discussion about Eqs.~\eqref{epsilonn}, we have,
\begin{equation}\label{beatmass1}
    \delta p=c_n\delta\lambda^{\delta_n+1}.
\end{equation}
Put it into Eq.~\eqref{massC3}, we have
\begin{equation}\label{massgb1}
\begin{split}
    2-\alpha_n^+(\lambda_0)&=\beta_n^+(\lambda_0+\delta\lambda)(\delta_n+1)+\\ &\frac{\ln[a^+_n(\lambda_0+\delta\lambda)|c_n|^{\beta^+_n(\lambda_0+\delta\lambda)}/b^+_n(\lambda_0)]}{\ln\delta\lambda}.
    \end{split}
\end{equation}
Then combining this with scaling laws \eqref{scalingrel}, we have,
\begin{equation}\label{massgb2}
    \frac{d\beta_n^+(\lambda_0)}{d\lambda_0}=-\lim_{\delta\lambda\rightarrow0}\frac{\ln[a^+_n(\lambda_0+\delta\lambda)|c_n|^{\beta^+_n(\lambda_0 +\delta\lambda)}/b^+_n(\lambda_0)]}{(\delta_n+1)\delta\lambda\ln\delta\lambda}
\end{equation}
As the critical exponents are independent of $\lambda$ in the neighbourhood of $\lambda_0$, so the right hand side of Eq. \eqref{massgb2} should be zero, which implies  Eq. \eqref{abc0}.

Note that our conditions for \eqref{abc0} don't mean that we need the critical exponents to be universal for all kinds initial value families. Figuratively speaking,   Eq. \eqref{abc0} holds  if there exists locally  at least one such a curve $p=p_n(\lambda)$ passing $A$ in Fig. \ref{plambda1}.

\section{Numerical check}
To confirm our scaling laws and the coefficient equation presented in Eqs. \eqref{scalingrel},  and \eqref{abc0}, here we show some numerical results for the critical exponents and proportionality coefficients.  As the value of $z_n^\pm$ can't be computed independently at present, we here only check the first scaling law in Eqs. \eqref{scalingrel} and the
coefficient equation  Eq. \eqref{abc0}. In the case with $p\rightarrow p_n^+$, we will use the model and algorithm presented in Ref.~\cite{Garfinkle}, which solved the massless scalar collapse in double null coordinates,
\begin{equation}\label{globm}
    ds^2=-fr'dudv+r^2d\Omega^2.
\end{equation}
Here $u$ and $v$ are two null coordinates, $r$ is the function of $u, v$, and a prime stands for the derivative with respect to
$v$. The initial value family we are considering is taken as,
\begin{equation}\label{initialv}
    \phi(0,v)=\left\{
    \begin{split}
    p &v^k\exp\left[-\lambda^{-2}\tan^2(\frac{v\pi}{2R})\right],~v<R\\
    &0,~v\geq R.
    \end{split}
    \right.
\end{equation}
with two tuning parameters $p$ and $\lambda$. In the following, we set $R=1$ and $\lambda_0=1/4$. A numerical check for the case of $p\rightarrow p^+_0$ is shown in Tab.~\ref{Tab1}.
\begin{table}
  \centering
  \begin{tabular}{c|cccccccc}
    \hline
    \hline
    $k$& $p_0$&$\beta_0^+$&$2-\alpha_0^+$&$\delta_0$&$a_0^+$&$b_0^+$&$|c_0|$&$|\Delta|$\\
    \hline
    3&84.3390&0.374&0.377&7E-4&1.81E-2&0.248&1.02E3&0.025\\
    4&365.775&0.376&0.377&2E-3&9.86E-3&0.260&5.85E3&0.011\\
    5&1465.05&0.376&0.378&-2E-3&5.58E-3&0.265&2.77E4&0.014\\
    \hline
  \end{tabular}
  \caption{The best fitting values about critical exponents and proportionality coefficients. Here $\Delta=1-a^+_0|c_0|^{\beta^+_0}/b^+_0$. }\label{Tab1}
\end{table}

In the case $p\rightarrow p^-_0$, we consider the gravitational collapse of a massless real scalar field in AdS space. In this case, a high precision simulation was done in Ref. \cite{Olivan:2015fmy}. We treat $\epsilon$ as $p$ and $\sigma$ as $\lambda$. Through the interpolation based on the data shown in Ref. \cite{Olivan:2015fmy}, in  the case $p_{00}=250, \lambda_0\approx0.0617$,  we  find $\beta_0^-\approx2-\alpha_0^-\approx0.7$,  $\delta_0\approx0$, and $a_0^-=2.29\times10^{-4}$, $b_0^-=8.46\times10^{-2}$, $c_0=4.81\times10^{3}$. Then we obtain $a^-_0|c_0|^{\beta^-_0}/b^-_0\approx1.02$. These results show that the first one in scaling laws \eqref{scalingrel} and the relation between the proportionality coefficients in Eq. \eqref{abc0} indeed hold up to numerical errors.

As for the anomaly dimension $z_n^{\pm}$, we find that $\beta_n^+\approx\nu_n^+$ for the metric ansatz \eqref{globm} and initial data \eqref{initialv}, which implies that $z_n^+=0$.  In the case with a gap studied in~\cite{Olivan:2015fmy}, due to the lack of precise data, we cannot get the anomaly dimension at the moment.

\section{Discussions}
\label{disc}
In our discussions, what we need are that the critical solution exists and the homogenous conditions \eqref{homo1} and \eqref{homo1t} hold. For a certain model,   the scaling transformations \eqref{homo1} and \eqref{homo1t} should be obtained from the model. In the case without the additional parameter $\lambda$ and in the asymptotic flat space-time, the value of $\beta_n^+$ can be obtained through the self-similarity near the original point and the analysis of Lyapunov exponent \cite{Koike:1995jm} when black hole nearly forms in the critical solution. However, how to use such method to obtain our scaling relations is still unknown. In more general, our knowledge about the critical behavior of gravitational collapse in the case of  $p\rightarrow p^-_n$ is still poor and the theoretical method beyond scaling analysis is an open question.

Though our discussions are concerned with the gravitational collapse, the scaling law \eqref{scalingrel} is more fundamental than what we have discussed. It is very interesting to compare our scaling laws with those  in thermodynamic criticality such as ferromagnetic phase transition or geometrical criticality such as percolation threshold \cite{Stauffer,Christensen}. By redefining the anomalous dimension $z_n^\pm$, one can find our scaling laws share the same forms in such two different systems. Such agreement is mysterious and needs to be understood further  in the future.

\section*{Acknowledgements}
This work was supported in part by the National Natural Science Foundation of China (No.11375247 and No.11435006).




\begin{thebibliography}{99}
\bibitem{Choptuik}
M. W. Choptuik,
``Universality and Scaling in Gravitational Collapse of a Massless Scalar Field,"
Phys. Rev. Lett. {\bf 70}, 9 (1993).

\bibitem{Gundlach:2007gc}
  C.~Gundlach and J.~M.~Martin-Garcia,
  ``Critical phenomena in gravitational collapse,''
  Living Rev.\ Rel.\  {\bf 10}, 5 (2007)
  doi:10.12942/lrr-2007-5
  [arXiv:0711.4620 [gr-qc]].

\bibitem{Hod:1996ar}
  S.~Hod and T.~Piran,
  ``Critical behavior and universality in gravitational collapse of a charged scalar field,''
  Phys.\ Rev.\ D {\bf 55}, 3485 (1997)
  [gr-qc/9606093].

\bibitem{Maison:1995cc}
  D.~Maison,
  ``Nonuniversality of critical behavior in spherically symmetric gravitational collapse,''
  Phys.\ Lett.\ B {\bf 366}, 82 (1996)
  [gr-qc/9504008].

\bibitem{Bizon:2011gg}
  P.~Bizon and A.~Rostworowski,
  ``On weakly turbulent instability of anti-de Sitter space,''
  Phys.\ Rev.\ Lett.\  {\bf 107}, 031102 (2011)
  [arXiv:1104.3702 [gr-qc]].

\bibitem{Maliborski:2012gx}
  M.~Maliborski,
  ``Instability of Flat Space Enclosed in a Cavity,''
  Phys.\ Rev.\ Lett.\  {\bf 109}, 221101 (2012)
  [arXiv:1208.2934 [gr-qc]].

\bibitem{Olivan:2015fmy}
  D.~S.~Oliv\'{a}n and C.~F.~Sopuerta,
  ``New features of gravitational collapse in Anti-de Sitter spacetimes,''
  arXiv:1511.04344 [gr-qc].

\bibitem{Cai:2015jbs}
  R.~G.~Cai, L.~W.~Ji and R.~Q.~Yang,
  ``Collapse of self-interacting scalar field in anti-de Sitter space,''
  arXiv:1511.00868 [gr-qc].

\bibitem{Garfinkle}
D. Garfinkle,
``Choptuik scaling in null coordinates,"
Phys. Rev. D {\bf 51}, 5558 (1995).

\bibitem{Koike:1995jm}
  T.~Koike, T.~Hara and S.~Adachi,
  ``Critical behavior in gravitational collapse of radiation fluid: A Renormalization group (linear perturbation) analysis,''
  Phys.\ Rev.\ Lett.\  {\bf 74}, 5170 (1995)
  [arXiv: gr-qc/9503007].

\bibitem{Stauffer}
D. Stauffer and A. Aharony,
 {\it Introduction to Percolation Theory}, 2nd edition (Taylor\&Francis, 1994).

\bibitem{Christensen}
 Kim Christensen, Nicholas R. Moloney,
{\it Complexity And Criticality (Imperial College Press Advanced Physics Texts)} (Imperial College Press, 2005),  Chapter 1.

\end{thebibliography}
\end{document}